# A Critical History of Renormalization[1]

Kerson Huang

Massachusetts Institute of Technology,
Cambridge MA, USA 02139
and
Institute of Advanced Studies, Nanyang Technological University
Singapore 639673

## Abstract

The history of renormalization is reviewed with a critical eye, starting with Lorentz's theory of radiation damping, through perturbative QED with Dyson, Gell-Mann & Low, and others, to Wilson's formulation and Polchinski's functional equation, and applications to "triviality", and dark energy in cosmology.

## 1. Dedication

   Renormalization, that astounding mathematical trick that enabled one to tame divergences in Feynman diagrams, led to the triumph of quantum electrodynamics. Ken Wilson made it physics, by uncovering its deep connection with scale transformations. The idea that scale determines the perception of the world seems obvious. When one examines an oil painting, for example, what one sees depends on the resolution of the instrument one uses for the examination. At resolutions of the naked eye, one sees art, perhaps, but upon greater and greater magnifications, one sees pigments, then molecules and atoms, and so forth. What is non-trivial is to formulate this mathematically, as a physical theory, and this is what Ken Wilson had achieved. To remember him, I recall some events at the beginning of his physics career.
   I first met Ken around 1957, when I was a fresh assistant professor at M.I.T., and Ken a Junior Fellow at Harvard's Society of Fellows. He had just gotten his Ph.D. from Cal. Tech. under Gell-Mann's supervision. In his thesis, he obtained exact solutions of the Low equation, which describes π-meson scattering from a fixed-source nucleus. (He described himself as an "aficionado" of the equation.) I had occasion to refer to this thesis years later, when Francis Low and I proved that the equation does not possess the kind of "bootstrap solution" that Geoffrey Chew advocated [2,3].

---

[1] To the memory of Kenneth G. Wilson (1936--2013)

While at the Society of Fellows, Ken spent most of his time at M.I.T. using the computing facilities. He was frequently seen dashing about with stacks of IBM punched cards used then for Fortran programming.

He used to play the oboe in those days, and I played the violin, and we had talked about getting together to play the Bach concerto for oboe and violin with a pianist (for we dare not contemplate an orchestra), but we never got around to that. I had him over for dinner at our apartment on Wendell Street in Cambridge, and received a thank-you postcard a few days later, with an itemized list of the dishes he liked.

At the time, my M.I.T. colleague Ken Johnson was working on non-perturbative QED, on which Ken Wilson had strong opinions. One day, when Francis Low and I went by Johnson's office to pick him up for lunch, we found the two of them in violent argument at the blackboard. So Francis said, "We'll go to lunch, and leave you two scorpions to fight it out."

That was quite a while ago, and Ken went on to do great things, including the theory of renormalization that earned him the Nobel Prize of 1982. In this article, I attempt to put myself in the role of a "physics critic" on this subject. I will concentrate on ideas, and refer technical details to [4,5].

While Ken's work has a strong impact on the theory of critical phenomena, I concentrate here on particle physics.

## 2. Lorentz: electron self-force and radiation damping

After J.J. Thomson discovered the first elementary particle, the electron [6], the question naturally arose about what it was made of. Lorentz ventured into the subject by regarding the electron as a uniform charge distribution of radius a, held together by unknown forces. As indicated in Fig.1, the charge elements of this distribution exert Coulomb forces on each other, but they do not cancel out, due to retardation. Thus, there is a net "self-force", and Lorentz obtained it in the limit $a \to 0$ [7]:

$$F_{\text{self}} = -m_{\text{self}} \ddot{x} + \frac{2e^2}{3c^3} \dddot{x} + O(a) \tag{1}$$

Internal Coulomb interactions give rise to a "self-mass":

$$m_{\text{self}} c^2 = \iint \frac{dq\,dq'}{|r - r'|} \xrightarrow[a \to 0]{} O\left(\tfrac{1}{a}\right) \tag{2}$$

which diverges linearly when $a \to 0$. This was the first occurrence of the "ultraviolet catastrophe", which befalls anyone toying with the inner structure of elementary particles.

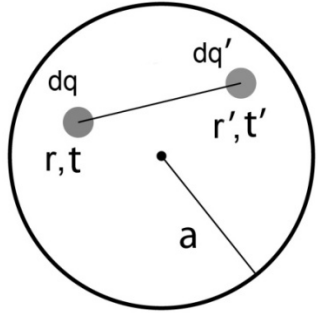

Fig1. Modeling the classical electron as charge distribution of radius a. The Coulombic forces between charge elements do not add up to zero, because of retardation: $|r-r'|=c|t-t'|$. Consequently, there is a "self-force", featuring a "self-mass" that diverges in the limit $a \to 0$, but can be absorbed into the physical mass. The finite remainder gives the force of radiation damping.

One notices with great relieve that the self-mass can be absorbed into the physical mass in the equation of motion

$$m_0 \ddot{x} = F_{self} + F_{ext}$$
$$\left(m_0 + m_{self}\right)\ddot{x} = \frac{2e^2}{3} \dddot{x} + F_{ext} + O(a) \qquad (3)$$

where $m_o$ is the bare mass. One can takes the physical mass from experiments, and write

$$m\ddot{x} = \frac{2e^2}{3c^3} \dddot{x} + F_{ext} \qquad (4)$$

with m=m₀+m_self. One imagines that the divergence of m_self is cancelled by m₀, which comes from the unknown forces that hold the electron together. This is the earliest example of "mass renormalization". Thus, the x term, the famous radiation damping, is exact in the limit $a \to 0$ within the classical theory. Of course, when a approaches the electron Compton wavelength, this model must be replaced by a quantum-mechanical one, and this leads us to QED (quantum electrodynamics).

# 3. The triumph of QED

Modern QED took shape soon after the advent of the Dirac equation in 1928 [8], and the hole theory in 1930 [9]. These theories make the vacuum a dynamical medium containing virtual electron-positron pairs. Weisskopf [10] was the first to investigate the electron self-energy in this light, and found that screening by induced pairs reduces the linear divergence in the Lorentz theory to a logarithmic one[2] [11,12]. Heisenberg, Dirac, and others [13-16] studied the electron's charge distribution due to "vacuum polarization", i.e., momentary charge separation in the Dirac vacuum. The unscreened "bare charge" was found to be divergent, again logarithmically. A sketch of the charge distribution of the electron is shown in Fig.2. The mildness of the logarithmic divergence played an important role in the subsequent renormalization of QED. But it was delayed for a decade because of World War II.

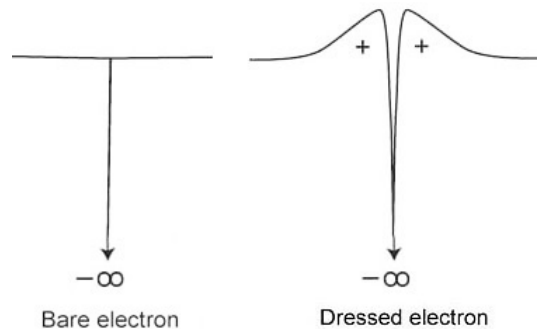

Fig2. Charge density of the bare electron (left) and that of the physical electron, which is "dressed" by virtual pairs induced in the Dirac vacuum (vacuum polarization). The bare charge is logarithmically divergent.

The breakthrough in QED came in 1947, with the measurements of the Lamb shift [17] and the electron anomalous moment [18]. In the first post-war physics conference at Shelter Island, LI, NY, June 2-4, 1947, participants thrashed out QED issues. (Fig.3 shows a group picture.) Bethe [19] made an estimate of the Lamb shift immediately after the conference, (reportedly on the train back to Ithaca, N.Y.,) by implementing charge renormalization, in addition to Lorentz's mass renormalization. This pointed the way to the successful calculation of the Lamb shift [17--19] in lowest-order perturbation theory.

---

[2] Weisskopf was then Pauli's assistant. According to his recollection (private communication), he made an error in his first paper and got a quadratic divergence. On day, he got a letter from "an obscure physicist at Harvard" by the name of Wendell Furry, who pointed out that the divergence should have been logarithmic. Greatly distressed, Weisskopf showed Pauli the letter, and asked whether he should "quit physics". The usually acerbic Pauli became quite restraint at moments like this, and merely huffed, "I never make mistakes!"

As for the electron anomalous moment, Schwinger [23] calculated it to lowest order as $\alpha/2\pi$, where $\alpha$ is the fine-structure constant, without encountering divergences.

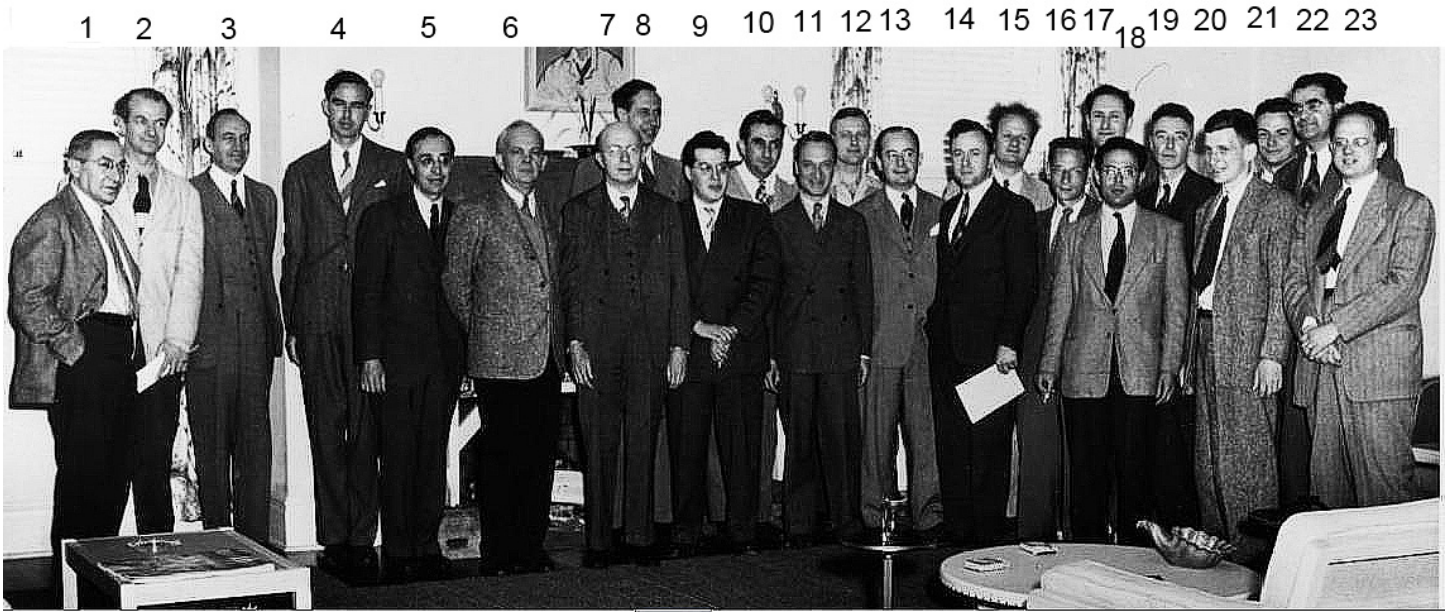

Fig.3  Shelter Island Conference on QED, June 2-4, 1947. Participants: 1 I. Rabi, 2 L. Pauling, 3 J. Van Vleck , 4 W. Lamb, 5 G. Breit, 6 Local Sponsor, 7 K. Darrow, 8 G. Uhlenbeck, 9 J. Schwinger, 10 E. Teller, 11 B. Rossi, 12 A. Nordsiek, 13 J. v. Neumann, 14 J. Wheeler, 15 H. Bethe, 16 R. Serber, 17 R. Marshak, 18 A. Pais, 19 J. Oppenheimer, 20 D. Bohm, 21 R. Feynman, 22 V. Weisskopf, 23 H. Feshbach.

## 4. Dyson: subtraction=multiplication, or the magic of perturbative renormalization

Dyson made a systematic study of  renormalization in QED in perturbation theory [24]. The dynamics in QED can be described in terms of scattering processes. In perturbation theory, one expands the scattering amplitude as power series in the electron bare charge $e_o$, (the charge that appears in the Lagrangian). Terms in this expansion are associated with Feynman graphs, which involve momentum-space integrals that diverge at the upper limit. To work with them, one introduces a high-momentum cutoff $\Lambda$. Dyson shows that showing that mass and charge renormalization remove all divergences, to all orders of perturbation theory.

The divergences can be traced to one of three basic divergent elements in Feynman graphs, contained in the full electron propagator S′ , the full photon propagator D′ , and the full vertex Γ. They can be reduced to the following forms:

$$S'(p) = [(p \cdot \gamma) - m_0 + \Sigma(p)]^{-1}$$
$$D'(k^2) = -k^{-2}[1 - e_0^2 \Pi(k^2)]^{-1}$$
$$\Gamma(p_1,p_2) = \gamma + \Lambda^*(p_1,p_2)$$

(5)

which must be regarded as power series expansions in $e_0^2$. The divergent elements are Σ,Π,Λ*, called respectively the self-energy, the vacuum polarization, and the proper vertex part. The Feynman graphs for these quantities are shown in Fig.4, and they are all logarithmically divergent. Thus, one subtraction will suffice to render them finite.

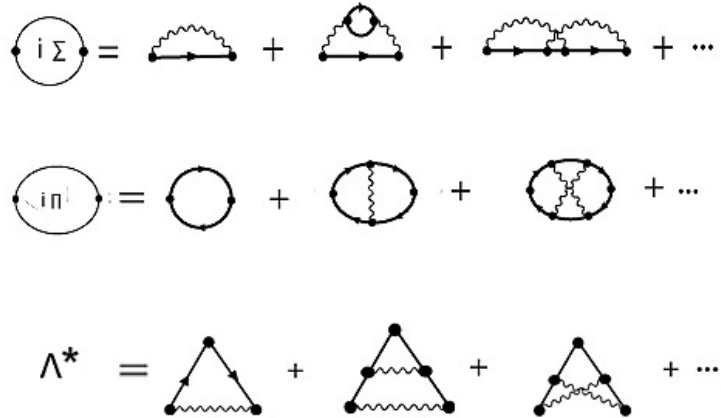

Fig4. The basic divergent elements in Feynman graphs. They are all logarithmically divergent.

The divergent part of Σ can absorbed into the bare mass $m_0$, as in the Lorentz theory. What is new is the divergent subtracted part of Π can be converted into multiplicative charge renormalization, whereby $e_0$ is replaced by the renormalized charge $e = Z e_0$. The divergence in Λ* can be similarly disposed of. We illustrate how this happen to lowest order.

The electron charge can be defined via the electron-electron scattering amplitude, which is given in QED by the Feynman graphs in Fig.5. The two electrons exchange a photon. We can write the propagator in the form

$$D'(k^2) = -\frac{d'(k^2)}{k^2}$$
$$d'(k^2) = \frac{1}{1 - e_0^2 \Pi(k^2)}$$

(6)

where k is the 4-momentum transfer.

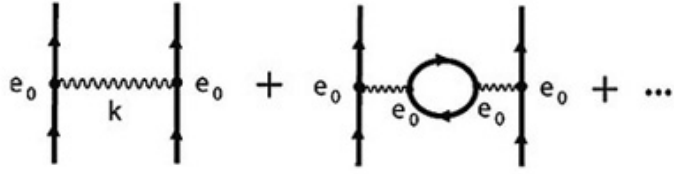

Fig5. Electron-electron scattering. The limit of zero 4-momentum transfer k→0 defines the electron charge.

To lowest order, the vacuum polarization is given by

$$\Pi(k^2) = -\frac{1}{12\pi^2} \ln \frac{\Lambda}{m} + R(k^2) + O(e_0^2) \tag{7}$$

where $\Lambda$ is the high-momentum cutoff, and m is the electron mass. (To this order it does not matter whether it is the bare mass or renormalized mass.) The first term is logarithmically divergent when $\Lambda \to \infty$, and the term R is convergent. One subtraction at some momentum μ makes Π convergent:

$$\Pi_c(k^2) \equiv \Pi(k^2) - \Pi(\mu^2) \tag{8}$$

We now write

$$e_0^2 d'(k^2) = \frac{e_0^2}{Z^{-1}(\mu^2) - e_0^2 \Pi_c(k^2)} = \frac{e_0^2 Z(\mu^2)}{1 - e_0^2 Z(\mu^2) \Pi_c(k^2)} \tag{9}$$

where

$$Z^{-1}(\mu^2) \equiv 1 - e_0^2 \Pi(\mu^2) \tag{10}$$

Both Z and $Z^{-1}$ are power series with divergent coefficients, and both diverge when $\Lambda \to \infty$. The combination $e_0^2 Z(\mu^2)$ gives a renormalized fine-structure constant

$$\alpha(\mu^2) = e_0^2 Z(\mu^2) \tag{11}$$

and the physical fine-structure constant corresponds to zero momentum transfer:

$$\alpha \equiv \alpha(0) \approx (137.036)^{-1} \tag{12}$$

We see that the subtraction of Π(μ²) in (8) has been turned into a multiplication by Z(μ²) in (11); but only to order $e_0^4$ in perturbation theory. Dyson proves the seeming miracle, that this is valid order by order, to all orders of perturbation theory.

## 5. Gell-Mann & Low: it's all a matter of scale

Gell-Mann and Low [25] reformulates Dyson's renormalization program, using a functional approach, in which the divergent elements Σ,Π,Λ* are regarded as functionals of one other, and functional equations for them can be derived from general properties of Feynman graphs. The divergent parts of these functionals can be isolated via subtractions, and the subtracted parts can be absorbed into multiplicative renormalization constant, by virtue of the behaviors of the functionals under scale transformations.

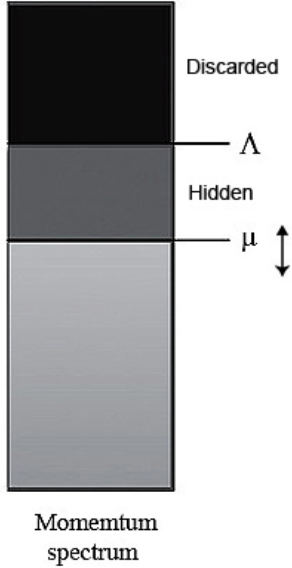

Fig6. The degrees of freedom of the system at higher momenta than the cutoff Λ are omitted from the theory, by definition. The degrees of freedom between Λ and the sliding renormalization point μ are "hidden" in the renormalization constants. Thus, μ is an effective cutoff, representing the scale at which one is observing the system.

One sees the cutoff Λ in a new light, as a scale parameter. In fact, it is the only scale parameter in a self-contained theory. When one performs a subtraction at momentum μ, and absorbs the Λ-dependent part into renormalization constants, one effectively lowers the scale from Λ to μ. The degrees of freedom between Λ and μ are not discarded, but hidden in the renormalization constants; the identity of the theory is preserved. The situation is illustrated in Fig.6.

The renormalized charge to order α², and for $|k|^2 \gg m^2$, is given by

$$\alpha(k^2) = \alpha + \frac{\alpha^2}{3\pi} \ln \frac{|k|^2}{m^2} \tag{13}$$

This is called a "running coupling constant", because it depends on the momentum scale k. It has been measured at a high momentum [26]:

$$\alpha(k_0^2) \approx (127.944)^{-1} \tag{14}$$

where $k_0 \approx 91.2$ GeV. The Fourier transform of $\alpha(k^2)$ gives the electrostatic potential of an electron [27]. As expected, it approaches the Coulomb potential $er^{-1}$ as $r \to \infty$, where e is the physical charge. For $r \ll \hbar/mc$, it is given by

$$V(r) \approx \frac{e}{r}\left[1 + \frac{2\alpha}{3\pi}\ln\frac{r_0}{r} + O(\alpha^2)\right] \tag{15}$$

where $r_0 = (\hbar/mc)(e^{5/6}\gamma)^{-1}$, $\gamma \approx 1.781$. We see the the bare charge $e_0$ of the electron, namely that residing at the center, diverges like $\ln(1/r)$.

Gell-Mann and Low [25] give the following physical interpretation of charge renormalization:

> A test body of "bare charge" $q_0$ polarizes the vacuum, surrounding itself by a neutral cloud of electrons and positrons; some of these, with a net charge δq, of the same sign as $q_0$, escape to infinity, leaving a net charge -δq in the part of the cloud which is closely bound to the test body (within a distance of $\hbar/mc$). If we observe the body from a distance much greater than $\hbar/mc$, we see an effective charge $q = q_0 - δq$, the renormalized charge. However, as we inspect more closely and penetrate through the cloud to the core of the test charge, the charge that we see inside approaches the bare charge $q_0$ concentrated at a point at the center.

## 6. Asymptotic freedom

The running coupling constant "runs" at a rate described by the β-function (introduced as ψ by Gell-Mann and Low):

$$\beta(\alpha(\mu^2)) = \mu^2 \frac{\partial \alpha(\mu^2)}{\partial \mu^2} \tag{16}$$

For QED we can calculate this from (13) to lowest order in α:

$$\beta_{\text{QED}}(\alpha) = \frac{\alpha^2}{3\pi} \tag{17}$$

That this is positive means that α increases with the momentum scale. But it has the opposite sign in QCD (quantum chromodynamics) [28,29]:

$$\beta_{\text{QCD}}(\alpha) = -\frac{\alpha^2}{6\pi}\left(\frac{33}{2} - N_f\right) \tag{18}$$

where α here is the analog of the fine-structure constant, and $N_f = 6$ is the number of quark flavors. Thus, QCD approaches a free theory in the high-momentum limit. This is called "asymptotic freedom".

QCD is a gauge theory like QED, but there are 8 "color" charges, and 8 gauge photons, called gluons, and. Unlike the photon, which is neutral, the gluons carry color charge. When a bare electron emits or absorbs a photon, its charge distribution does not change, because the photon is neutral. In contrast, when a quark emits or absorbs a gluon, its charge center is shifted, since the gluon is charged. Consequently, the "dressing" of a bare quark smears out its charge to a distribution without a central singularity. As one penetrates the cloud of vacuum polarization of a dressed quark, one see less and less charge inside, and finally nothing at the center. This is the physical origin of asymptotic freedom. Fig.7 shows a comparison between the dressed electron and the dressed quark, with relevant Feynman graphs that contribute to the dressing.

In the standard model of particle physics, there are 3 forces strong, electromagnetic and weak, whose strengths can be characterized respectively by $\alpha_{QCD}$, $\alpha_{QED}$, $\alpha_{Weak}$, with strength standing at low momenta in the approximate ratio 10: $10^{-2}$: $10^{-5}$. While $\alpha_{QCD}$ is asymptotically free, the other two are not. Consequently $\alpha_{QCD}$ will decrease with momentum scale, whereas the other two increase. Extrapolation of present trend indicate they would all meet at about $10^{17}$ GeV, as indicated in Fig.8. This underlies the search for a "grand unified theory" at that scale.

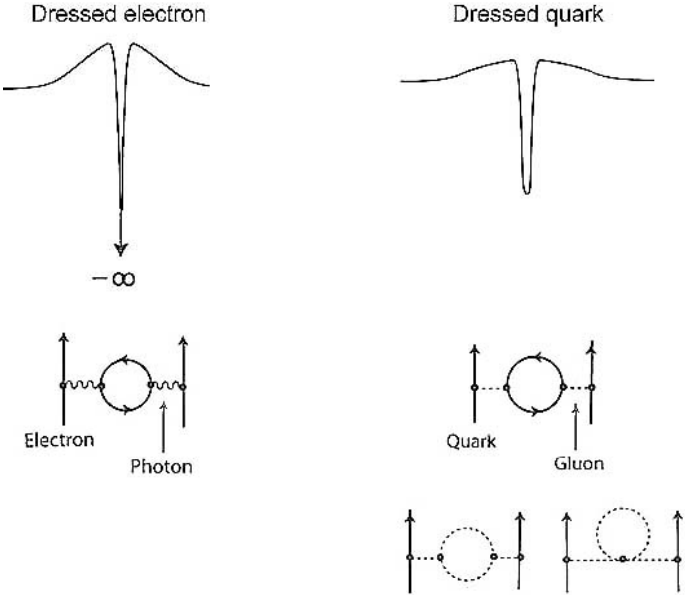

Fig7. Comparison between a dressed electron and a dressed quark. There is a point charge at the center of the dressed electron, but none in the dressed quark, for it has been smeared out by the gluons, which are themselves charged. Lower panels show the relevant Feynman graphs. For the quark, there are two extra graphs arising from gluon-gluon interactions.

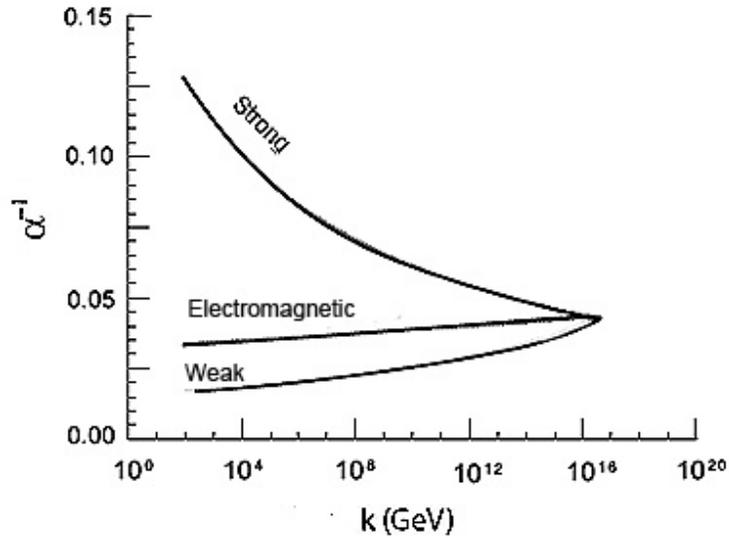

Fig8. Extrapolation of the running coupling constants for the strong, electromagnetic, and weak interactions indicate that they would meet at a momentum k≈ 10¹⁷GeV, giving rise to speculations of a "grand unification".

# 7. The renormalization group (RG)

The transformations of the scale μ form form a group, and the running coupling constant α(μ²) gives a representation of this group, which was named RG (the renormalization group) by Bogoliubov [30]. The β-function is a "tangent vector" to the group. By integrating (16), we obtain the relation

$$\ln\frac{\mu^2}{\mu_0^2} = \int_{\alpha(\mu_0^2)}^{\alpha(\mu^2)} \frac{d\alpha}{\beta(\alpha)}$$

(19)

As $\mu \to \infty$, the left side diverges, and therefore $\alpha(\mu^2)$ must either diverge, or approach a zero of $\beta(\alpha)$. The latter is a fixed point of RG, at which the system is scale-invariant. This shows that the scale $\mu$ is determined by the value of $\alpha(\mu^2)$.

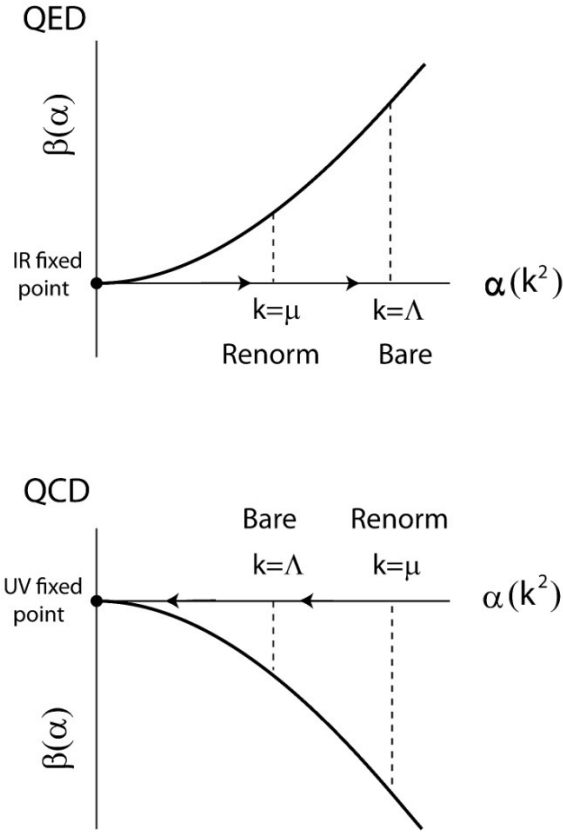

Fig9. The sign of β(α) determines the direction of arrows on the α axis, indicating the direction of running as the momentum scale k increases. Fixed points are where β(α)=0, at the system is scale-invariant, and. The origin is an IR (infrared) fixed point for QED, for the system goes towards it when k→0. It is an UV (ultraviolet) fixed point for QCD, which approaches it when k→∞.

Fig.9 shows plots of the β-function for QED and QCD. As the momentum scale k increases, α(k²) runs along the direction of the arrows determined by the sign of β. For QED, α increases with k, and since perturbation theory becomes invalid at high k, we lose control over high-energy QED. For QCD, on the other hand, α runs towards the UV fixed point at zero, perturbation theory becomes increasingly accurate, and we have a good understanding in this regime. The other side of the coin is that QCD becomes a hard problem at low energies, where it exhibits quark confinement.

The plots clarify the relation between the cutoff scale $\Lambda$ that defines the bare system, and the effective scale μ, which defines the renormalized system.
We now have a better understanding of what can be done with the original cutoff $\Lambda$. Being a scale parameter, $\Lambda$ is determined by (19) and the limit $\Lambda \to \infty$ can be achieved only by moving $\alpha(\Lambda^2)$ to a fixed point, and in QCD this means $\alpha(\Lambda^2)$=0. And there is no problem with this.

In QED, on the other hand, there is no known fixed point except the one at the origin. In practice, one keeps Λ finite, whose value is not important. In this way, one can perform calculations that agree with experiments to one part in $10^{12}$, in the case of the electron anomalous moment [29,30]. If one insists on making Λ infinite, one must make $\alpha(\Lambda^2)=0$, but that makes $\alpha(\mu^2)=0$ for all $\mu<\Lambda$, and one has a trivial free theory. We will expand on this "triviality problem" later.

Particle theorists have a peculiar sensitivity to the cutoff, because they regard it as a stigma that exposes an imperfect theory. In the early days of renormalization, when the cutoff was put out of sight by renormalization, some leaped to declare that the cutoff has been "sent to infinity". That, of course, cannot be done by fiat. Only in QCD can one achieve that, owing to asymptotic freedom.

A more general statement of renormalization refers to any correlation function G':

$$G'(p;\Lambda,e_0^2) = Z^*(\Lambda/\mu,e_0^2)G(p;\mu,\alpha(\mu^2))$$
(20)

where p collectively denotes all the external external momenta, Z* is a dimensionless renormalization constant that diverges when Λ→∞, μ is an arbitrary momentum scale less than Λ, α(μ²) is given by (11), and G is a convergent correlation function. Since the left side is independent of μ, we have

$$\frac{d}{d\mu}\left[Z^*\left(\Lambda/\mu,e_0^2\right)G(p;\mu,\alpha)\right]=0$$
(21)

which leads to the Callan-Symanzik equation [32,33]

$$\left[\mu\frac{\partial}{\partial\mu}+\beta(\alpha)\frac{\partial}{\partial\alpha}+\gamma(\mu)\right]G(p;\mu,\alpha)=0$$
(22)

where β is defined by (16), and γ(μ)=μ(∂/∂μ) lnZ*(Λ/μ, α₀) is called the "anomalous dimension". This shows how renormalization accompanies a scale transformation, so as to preserve the basic identity of the theory.

## 8. The Landau ghost

Between the great triumph of quantum field theory in QED in 1947, and the emergence of the standard model of particle physics around 1975, particle theorists wandered like Moses in some desert, for nearly three decades. During that time they get disenchanted with quantum field theory, because the great hope they had pinned on the theory to explain the strong interactions did not materialize[3]. The was a feeling that something crazy was called for, like quantum

---
[3] I recall that, in the late fifties, E. Fermi and F.J. Dyson separately gave the Morris Loeb Lecture at Harvard University. Fermi talked about a newly discovered pio-nucleon "33 resonance", with spin 3/2 and isospin 3/2 (now

mechanics[4], or maybe the "bootstrap" [2,3]. Landau thought he has at least disposed of quantum field theory by exposing a fatal flaw.

Substituting (16) into (19) and performing the integration, one obtains

$$\alpha(k^2) = \frac{\alpha}{1 - (\alpha/3\pi)\ln(k^2/m^2)}$$
(23)

This is supposed to be an improvement on (13), equivalent to summing a certain class of Feynman graphs --- the so-called "leading logs" with terms of the form $(e_0^2 \ln \Lambda)^n$. Landau [34] pointed out that there is a pole with negative residue:

$$\alpha(k^2) \approx \frac{-3\pi k^2_{ghost}}{k^2 - k^2_{ghost}}$$

$$\frac{k^2_{ghost}}{m^2} = \exp\frac{3\pi}{\alpha}$$
(24)

This represents a photon excited state, whose wave function has negative squared modulus, and is called a "ghost state". Its mass is of order $10^{300}$ m. It can be shown that $\Lambda < k_{ghost}$, and thus the ghost occurs only if we continue the theory to beyond the preset cutoff. However, if one insists on making $\Lambda \to \infty$, one must push the ghost to infinity, and this means $\alpha \to 0$, leading to a trivial theory. Landau said that this possibility exposes a fundamental flaw in quantum field theory[5], which "should be buried with honors".

The triviality problem also occurs in other theories, for example the scalar $\phi^4$ Higgs field in the standard model. Earlier, it was found in the Lee model [35], an exactly soluble model of meson scattering. Källén and W. Pauli [36] showed that

---

called the delta baryon), and said, "I will not understand this in my lifetime." Dyson talked about the so-called "Tamm-Dancoff approximation" for pion-nucleon scattering, and said, "We will not understand this problem in a hundred years."

[4] In 1958, Heisenberg and Pauli proposed a "unified field theory". Pauli gave a seminar at Columbia University with Niels Bohr in attendance. When the seminar began, Bohr said, "To be right, the theory had better be crazy". Pauli said, "It's crazy! You will see. It's crazy!" The theory turns out be a version of the four-fermion interaction.

[5] Apparently, Landau considered the ghost state a hallmark of quantum field theories. He reportedly calculated the β-function of Yang-Mills theory (on which QCD is based), but made a sign error, and missed asymptotic freedom.

the ghost state renders the S-matrix non-unitary, and this pathology cannot be cured by redefining Hilbert space to admit negative norms.[6]

We shall see that the triviality problem is a general property of IR fixed points. The moral is: to get infinite cutoff, get yourself a UV fixed point!

Quantum field theory did not die, but bounced back with a vengeance, in the form of Yang-Mills gauge theory in the standard model.

## 9. Renormalizability

Renormalization in perturbation theory hinges on the degree of divergence K of Feynman graphs, which is determined via a power-counting procedure. It depends on the form of coupling --- how many lines meet at a vertex, etc. Renormalization in QED relies on the fact the interaction $\bar{\psi}A\psi$ gives K=0 (logarithmic divergence). One can imagine interactions that would give K>0, and that would be non-renormalizable. An example is the 4-fermion interaction $(\bar{\psi}\psi)^2$. There is thus a criterion of renormalizability: under a scale change, the existing coupling constant undergo renormalization, and no new coupling should arise. In other words, the system should be self-similar.

Such considerations are based on the presumption that each new coupling bring in its own scale. In a self-contained system, however, the cutoff Λ sets the only scale, and all coupling constants must be proportional to an appropriate power of Λ. When this is taken into account in the power counting, what was considered a non-renormalizable interaction can become renormalizable. If all coupling constants are made dimensionless in this manner, then they could freely arise under scale transformations, and the system need not be self-similar to be renormalizable.

As illustration, consider scalar field theory with a Lagrangian density of the form (with ℏ=c=1)

$$\mathcal{L} = \frac{1}{2}(\partial \phi)^2 - V(\phi)$$
$$V(\phi) = g_2 \phi^2 + g_4 \phi^4 + g_6 \phi^6 + \cdots$$
(25)

The theory is called ϕ$^M$ theory, where M is the highest power that occurs. Each coupling $g_n$ corresponds to a vertex in a Feynman graph, at which n lines meet, and each line carries momentum. The momenta of the internal lines are integrated over, and produce divergences. Thus, each Feynman graph is proportional to Λ$^K$, with a degree of divergence K that can be found by a counting procedure. The relation between K and topological properties of

---

[6] There other ghost states in quantum field theory, arising from gauge-fixing, such the Fadeev-Popov ghost [5]. But these are mathematical devices that have no physical consequences.

Feynman graphs, such as the number of vertices and internal lines, determines renormalizability.

It was said conventionally that only the φ⁴ theory is renormalizable. This determination, however, assumes that the $g_n$ are arbitrary parameters. The dimensionality of $g_n$ in d-dimensional space-time is

$$[g_n] = (\text{Length})^{nd/2-n-d} \tag{26}$$

Treating them as independent will means that each $g_n$ bring into the system an independent scale. But the only intrinsic length scale in a self-contained system is the inverse cutoff $\Lambda^{-1}$. Thus each $g_n$ should be scaled with appropriate powers of $\Lambda$:

$$g_n = u_n \Lambda^{n+d-nd/2} \tag{27}$$

so that $u_n$ is dimensionless. When this is done, the cutoff dependence of $g_n$ enters into the power counting, and all $\phi^K$ theories become renormalizable [37].

With the scaling (27), one can construct an asymptotically free scalar field, one that is free from the triviality problem. For an N-component scalar field in d=1, V(φ) is uniquely given by the Halpern-Huang potential [38]

$$V(\phi) = c\Lambda^{4-b}[M(-2+b/2, N/2, z) - 1]$$

$$z = \frac{8\pi^2}{\Lambda^2} \sum_{n=1}^{N} \phi_n^2 \tag{28}$$

where c,b are arbitrary constants, and M(a,b;z) is the Kummer function, which has exponential behavior for large fields:

$$M(p,q,z) \approx \Gamma(q)\Gamma^{-1}(p)z^{p-q}\exp z \tag{29}$$

The theory is asymptotically free for b>0. This has applications in the Higgs sector of the standard model and in cosmology, to be discussed later.

Not all theories are renormalizable, even with the scaling of coupling constants. There is a true spoiler, namely, the "axial anomaly" in fermionic theories. It arises from the fact that the classically conserved axial vector current becomes non-conserved in quantum theory, due to the existence of topological charges. (See [2,4]). This leads to Feynman graphs with the "wrong" scaling behavior, and the only way to get rid of divergences arising from such Feynman graphs is to cancel them with similar graphs. The practical consequence is that quarks and leptons in the standard model must occur in a family, such that their anomalies cancel. We know of three families: {u,d,e,$v_e$}, {s,c,μ, $v_\mu$ }, {t,b,τ, $v_\tau$ }. If a new quark or lepton is discovered, it should bring with it an entire family.

# 10. Wilson's renormalization theory

   Wilson reformulates renormalization independent of perturbation theory, and puts scale transformations at the forefront. He was concerned with critical phenomena in matter, where there is a natural cutoff, the atomic lattice spacing a. When one writes down a Hamiltonian, a does not explicitly appear, because it only supplies the length scale. The scaling (27) of coupling constants is natural and automatic. This is an important psychological factor in one's approach to the subject.

   The first hint of how to do renormalization on a spatial lattice space comes from Kadanoff's "block spin" transformations [39]. This is a coarse-graining process, as illustrated in Fig.10. Spins with only up-down states are represented by the black dots, with nearest-neighbor (nn) interactions. In the first level of coarse-graining, spins are grouped into blocks, indicated by the solid enclosures. The original spins are replaced by a single averaged spin at the center. The lattice spacing becomes 2, but is rescaled back to 1. The block-block interactions now have renormalized coupling constants; however, new couplings arise, for the blocking process generates nnn and longer-ranged interactions. Kadanoff concentrates on the fixed points of iterative blocking, and ignores the new couplings for this purpose. Wilson take the new couplings into account, by providing "hooks" for them from the beginning. That is, the coupling-constant space is enlarged to include all possible couplings: nnn, nnnn, etc. In the beginning, when there were only the nn couplings, one regarded the rest as potentially present, but negligible. The couplings can grow or decrease in successive blocking transformations.

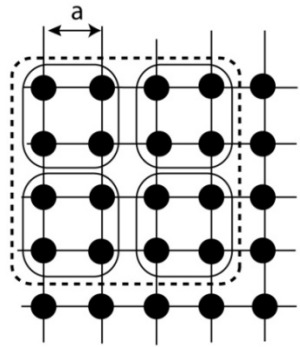

Fig10.   Block-spin transformations. In the spin lattice, the up-down spins, represented by the black dots, interact with each other via nearest-neighbor (nn) interactions. In the first level of coarse-graining, they are grouped into blocks of 4, indicated by the solid enclosures, and replaced by a single averaged spin at the center. The original lattice spacing a now becomes 2a, but is rescaled back to a. In the next level, these blocks are grouped into higher blocks indicated by the dotted enclosure, and so forth. However,the block-block interactions will include nnn, nnnn interactions, and so forth.

Wilson implements renormalization using the Feynman path integral, as follows. A quantum field theory can be described through its correlation functions. For a scalar field, for example, these are the functional averages <ϕϕ>, <ϕϕϕ>, <ϕϕϕϕ>, ..., and they can be obtained from the generating functional

$$W[J] = \mathcal{N} \int D\phi \exp i(S[\phi, \Lambda] - (J, \phi))$$

(30)

by repeated functional differentiation with respect to the external current *J(x)*. Here $S[\varphi] = \int d^d x \, \mathcal{L}$ is the classical action, where $\mathcal{L}$ is the classical Lagrangian density (25), and $\int D\phi$ denotes functional integration. There is a short-distance cutoff $\Lambda^{-1}$, which is only scale in S[ϕ]. Of course, *J* introduces an scale, but that is external rather than intrinsic. For simplicity we set J≡0 in this discussion.

By making the time pure imaginary (Euclidean time, in the language of relativistic quantum field theory) one can regard W[J] as the partition function for a thermal system described by an order parameter ϕ(x), and the imaginary time corresponds to the inverse temperature. In this way, a result from quantum field theory can be translated into that in statistical mechanics, and vice versa.

The functional integration $\int D\phi$ extends over all possible functional forms of ϕ(x). It may be carried out by discretizing x as a spatial lattice, and integrating over the field at each site. Alternatively one can integrate over all Fourier transforms in momentum space, made discrete by enclosing the system in a large spatial box. Here we choose the latter route:

$$\int D\phi = \prod_{|k|<\Lambda} \int_{-\infty}^{\infty} d\phi_k$$

(31)

where ϕ$_k$ denotes a Fourier component of the field, and Λ is the high-momentum cutoff. We lower the effective cutoff to μ by "hiding" the degrees of freedom between Λ and μ, as indicated in Fig.6. To do this, we integrate over the momenta in this interval, and put the result in the form of a new effective action. That is, we write

$$\mathcal{N} \prod_{|k|<\Lambda} \int_{-\infty}^{\infty} d\phi_k \exp iS[\phi] = \mathcal{N} \prod_{|k|<\mu} \int_{-\infty}^{\infty} d\phi_k \left\{ \prod_{\mu<|k|<\Lambda} \int_{-\infty}^{\infty} d\phi_{k'} \exp iS[\phi] \right\}$$

$$= \mathcal{N}' \prod_{|k|<\mu} \int_{-\infty}^{\infty} d\phi_k \exp iS'[\phi]$$

(32)

The integrations in the brackets $\{\}$ define the new action $S'[\phi]$, which contains only degrees of freedom below momentum μ [7]. From this, we can obtain a new Lagrangian density $\mathcal{L}'$, which contains new couplings $\{u'_n\}$ that are functions of the old ones $\{u_n\}$ [8]. This, in a nutshell, is Wilson's renormalization transformation.

---

[7] The cutoff Λ actually does not appear in any of the formulas, because it merely supplies a scale. Lowering the cutoff from Λ to μ actually means lowering it from 1 to μ/Λ. See [4] for details.

Successive renormalization transformations give a series of effective Lagrangians:

$$\mathcal{L} \to \mathcal{L}' \to \mathcal{L}'' \to \mathcal{L}''' \to \cdots \qquad (33)$$

which describe how the appearance of the system changes under coarse-graining. The identity of the system is preserved, because the generating functional W is not changed. We allow for all possible couplings $u_n$, and thus the parameter space is that of all possible Lagrangians. Renormalization generates a trajectory in that space --- the RG trajectory. Couplings that were originally negligible can grow, and so the trajectory can break out into new dimensions, as illustrated in Fig.11. There is no requirement that the theory be self-similar, and thus it appears that all theories are renormalizable[9].

That this method of renormalization reduces to that in perturbation theory can be proven by deriving (20) with this approach [4].

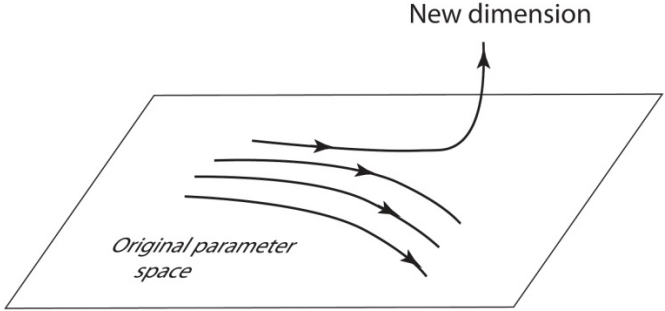

Fig11. By rendering all coupling constant dimensionless through scaling with appropriate powers of the cutoff momentum, the system can break out into a new direction in parameter space under renormalization. The trajectories sketched here represent RG trajectories with various initial conditions.

---

[8] Some rescaling need to be done to put the new Lagrangian in a standard form. These operations can affect the normalization constant N.

[9] Notable exceptions, as mentioned previously, are fermionic theories exhibiting the axial anomaly. In terms of the Feynman path integral, certain scaling operations fail, owing to non-invariance of the integration measure. See [2,4].

## 11. In the space of all possible Lagrangians

Under the coarse-graining steps, the effective Lagrangian traces out a trajectory in parameter space, the RG trajectory[10]. With different initial conditions, one goes on different trajectories, and the whole parameter is filled with them, like streams lines in a hydrodynamic flow. There are sources and sinks in the flow, and these are fixed points, where the system remains invariant under scale changes. The correlation length becomes infinite at these fixed points. This means that the lattice approaches a continuum: a→0, or $\Lambda \to \infty$.

Let us define the direction of flow along an RG trajectory to be the coarse-graining direction, or towards low momentum. If it flows out of a fixed point, then the fixed point appears to be a UV fixed point, for it is to be reached by going opposite to the flow, towards the high-momentum limit. Such a trajectory is called a UV trajectory. If if flows into a fixed point, it is called an IR trajectory, along which the fixed point appears to be an IR fixed point. This is illustrated in Fig.12.

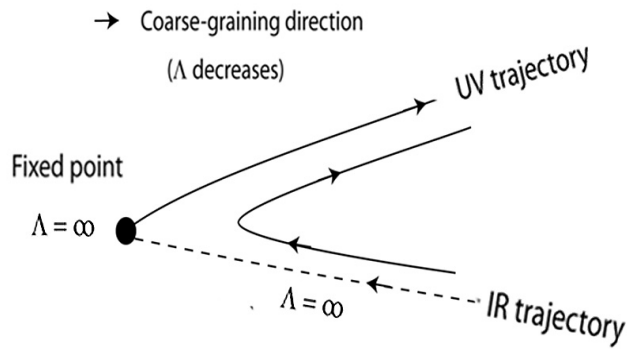

Fig12. Fix points are sources and sinks of RG trajectories, at which the cutoff is infinite.

Actually, Λ is infinite along the entire IR trajectory, because this is so at the fixed point, and Λ can only decrease upon coarse-graining. Thus, one cannot place a system on an IR trajectory, but only on an adjacent trajectory. When we get closer and closer to the IR trajectory, $\Lambda \to \infty$, and system more closely resemble that at the IR fixed point. It is most common to have a free theory at the fixed point, since it is scale-invariant, and this gives a more physical understanding of the triviality problem.

---

[10] The coarse-graining proceeds only in one direction; but that is a matter of defining the RG trajectory. Once defined, one can travel back and forth along the trajectory.

The flow velocity along an RG trajectory can be measured by the arc length covered in a coarse-graining step. It slows down in the neighborhood of a fixed point, and speed up between fixed points. Thus, it darts from fixed point to the next, like a ship sailing between ports of call. Some couplings grow as it approaches a fixed point, and these are called "relevant" interactions. The one that die out are called "irrelevant", and may be neglected. Thus, each port corresponds to a characteristic set of interactions, and the system puts on a certain face at that port. This is illustrated in Fig.13.

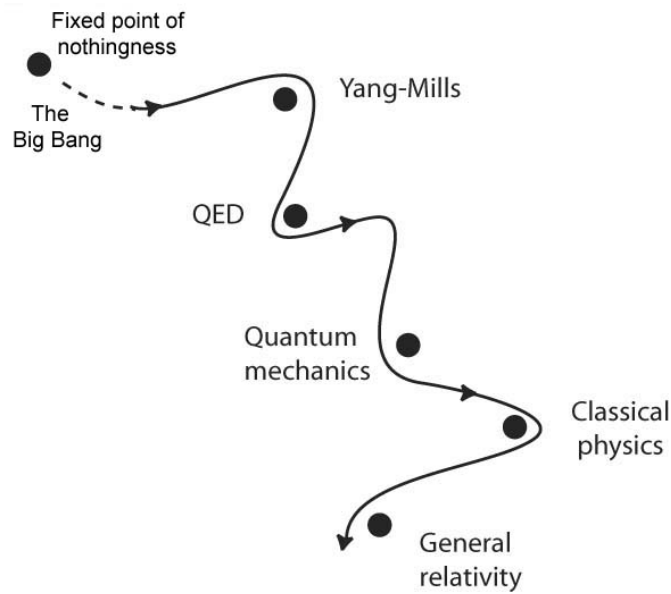

Fig13. Different physical theories govern different length scales. Each theory can be represented by a fixed point in the space of Lagrangians. The world is like a ship navigating this space, and the fixed points are the ports of call. As the scale changes, the world sails from port to port, and lingers for a while at each port.

## 12.  Polchinski's functional equation

We have used a sharp momentum cutoff in (31). In a significant improvement, Polchinski [41] generalizes the method to an arbitrary cutoff, and derive a functional equation for the renormalized action. The cutoff is introduced by modifying the free propagator $k^{-2}$ of the field theory, by replacing it by

$$\Delta(k^2) = \frac{f(k^2/\Lambda^2)}{k^2}$$

$$f(z) \xrightarrow[z\to\infty]{} 0 \tag{34}$$

where the detailed form of the cutoff function f(k²/Λ²) is not important. What is important is that Λ is the only scale in the theory. The regulated propagator in configurational space will be denoted by K(x,Λ).

The action is written as

$$S[\phi,\Lambda] = S_0[\phi,\Lambda] + S'[\phi,\Lambda] \tag{35}$$

where the first term corresponds to the free field, and the second term represents the interaction. We have

$$S_0[\phi,\Lambda] = \frac{1}{2}\int d^d x\, d^d y\, \phi(x) K^{-1}(x-y,\Lambda)\phi(y) \tag{36}$$

where K⁻¹(x-y,Λ) is the inverse of the propagator K(x-y,Λ), in an operator sense. It differs from the Laplacian operator significantly only in a neighborhood of |x-y|=0, of radius Λ⁻¹. The generating functional is written as

$$W[J,\Lambda] = \mathcal{N}\int D\phi\, e^{-S[\phi,\Lambda]-(J,\phi)} \tag{37}$$

where the normalization constant N may depend on Λ.

In the Wilson method, one integrates out mode between Λ and μ to lower the effective cutoff to μ. A more general point of view is that any change in Λ is compensated by a change in S'[Φ,Λ], in order to preserve the basic identity of the theory:

$$\frac{dW[J,\Lambda]}{d\Lambda} = 0 \tag{38}$$

This is the generalization of (21) in perturbative renormalization.

The remarkable thing is that Polchinski solves (38) by finding a functional integro-differential equation for S'[Φ,Λ]. For J≡0, it reads[11]

$$\frac{dS'}{d\Lambda} = -\frac{1}{2}\int dx\, dy\, \frac{\partial K(x-y,\Lambda)}{\partial \Lambda}\left[\frac{\delta^2 S'}{\delta\phi(x)\delta\phi(y)} - \frac{\delta S'}{\delta\phi(x)}\frac{\delta S'}{\delta\phi(y)}\right] \tag{39}$$

Periwal [42] shows how one can use this to derive the Halpern-Huang potential in "two lines". (The original derivation involves summing one-loop Feynman graphs [38]). Assuming that there are no derivative couplings[12], we can write S' as the integral of a local potential:

---

[11] See [4] for a proof.

[12] The original derivation did not assume this, but showed that no derivative couplings arise from renormalization, if none were present originally.

$$S'[\phi,\Lambda] = \Lambda^d \int d^d x\, U(\varphi(x),\Lambda)$$

$$\varphi(x) = \Lambda^{1-d/2}\phi(x) \tag{40}$$

where U is a dimensionless function, and φ is a dimensionless field. The scalar potential is given by $V=\Lambda^d U$.

Near the Gaussian fixed point, where S'≡0, one can linearize (39), and obtain a linear differential equation for U(φ,Λ):

$$\Lambda\frac{\partial U}{\partial \Lambda} + \frac{\kappa}{2}U'' + \left(1-\frac{d}{2}\right)\varphi U' + Ud = 0 \tag{41}$$

where a prime denote partial derivative with respect to φ, and $\kappa = \Lambda^{3-d}\partial K(0,\Lambda)/\partial \Lambda$. Now we seek eigenpotentials $U_b(\varphi,\Lambda)$ with the property

$$\Lambda\frac{\partial U_b}{\partial \Lambda} = -bU_b \tag{42}$$

In the language of perturbative renormalization theory, the right side is the linear approximation to the β-function., and the solution is asymptotically free for b>0. Substitution into the previous equation leads to the differential equation

$$\left[\frac{\kappa}{2}\frac{d^2}{d\varphi^2} - \frac{1}{2}(d-2)\varphi\frac{d}{d\varphi} + (d-b)\right]U_b = 0 \tag{43}$$

Since this equation does not depend on Λ, the Λ-dependence of the potential is contained in a multiplicative factor. In view of (42), the factor is $\Lambda^{-b}$.

For d≠2, (42) can be transformed into Kummer's equation:

$$\left[z\frac{d^2}{dz^2} + (q-z)\frac{d}{dz} - p\right]U_b = 0 \tag{44}$$

where

$$q = 1/2$$
$$p = \frac{b-d}{d-2}$$
$$z = (2\kappa)^{-1}(d-2)\varphi^2 \tag{45}$$

The solution is

$$U_b(z) = c\Lambda^{-b}[M(p,q,z) - 1] \tag{46}$$

where c is an arbitrary constant, and M is the Kummer function. We have subtracted 1 to make $U_b(0)=0$. This is permissible, since it merely changes the normalization of the generating functional. For d=2, (43) leads to the so-called sine-Gordon theory.

## 13. Why triviality is not a problem

The massless free scalar is scale-invariant, and corresponds to the Gaussian fixed point. When the length scale increases from zero, and we imagine the system being displaced infinitesimally from this fixed point, it will sail along some RG trajectory, along some direction in parameter space, the function space spanned by possible forms of $V(\phi)$. Eq.(42) describes the properties associated with various directions. Along directions with b>0, the system will be on a UV trajectory. With b<0, the system is on an IR trajectory, and behave as if it had never left the fixed point. This is illustrated in Fig.14.

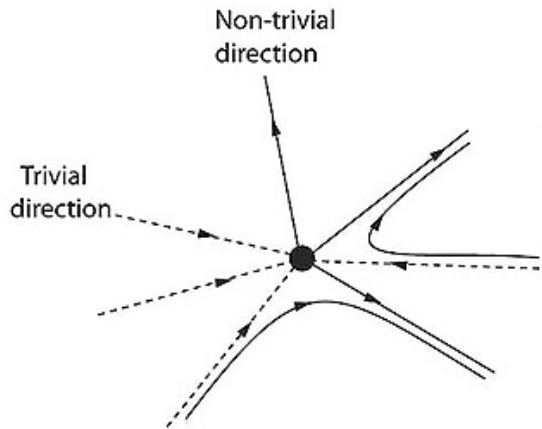

Fig14. The Gaussian fixed point. RJ trajectories emanate along all possible directions in parameter space. Arrows denote direction of increasing length scale. Non-trivial directions correspond to theories with asymptotic freedom. Trivial direction signifies that the theory remains at the fixed point under scale change.

The existence of UV directions suggests a possible solution to the triviality problem, as illustrated in Fig.15. Consider two Gaussian fixed points A and B. A scalar field leaves A along a UV trajectory, and crosses over to a neighborhood B, skirting an IR trajectory of B. At point 1, the potential is Halpern-Huang, but at 2 it becomes $\phi^4$, with all higher coupling becoming irrelevant. The original cutoff Λ is infinite, being pushed into A. The effective cutoff at 2 is a renormalization point μ, and there is no reason to make it infinite.

In the case of QED, the fixed point B would correspond to our QED Lagrangian, and A could represent some asymptotically free Yang-Mill gauge theory.

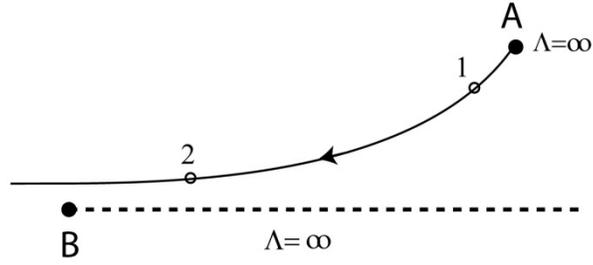

Fig15. How "triviality" may arise, and why it is not a problem. Here, A and B represent two Gaussian fixed points. The system at 1 is on a UV trajectory and asymptotically free. It crosses over to a neighborhood of B, skirting an IR trajectory. At point 2 it resembles a trivial φ⁴ theory, because higher couplings have become irrelevant.

## 14. Asymptotic freedom and the big bang

The vacuum carries complex scalar fields. There is at least the Higgs field of the standard model, which generates mass for gauge bosons in the weak sector. Grand unified theories call for more scalar fields. A complex scalar field serves as order parameter for superfluidity, and from this point of view the entire universe is a superfluid. In a recent theory, dark energy and dark matter in the universe arise from this superfluid. Briefly, dark energy is the energy density of the superfluid, and dark matter is the manifestation of density fluctuations of the superfluid from its equilibrium vacuum value [43-45].

At the big bang, the scalar field is assumed to emerge from the Gaussian fixed point along some direction in parameter space, as indicated in Fig.14. If the chosen direction corresponds to an IR trajectory, then the system never left the fixed point, and nothing happens. If it is a UV trajectory, however, it will develop into a Halpern-Huang potential, and spawn a possible universe. We assume that was only one scale at the big bang, the radius of the universe a(t) in the Robertson-Walker metric. Thus, it must be identified with the cutoff Λ of the scalar field:

$$\Lambda = \frac{\hbar}{a(t)} \qquad (47)$$

This relation creates a dynamical feedback: the scalar field generates gravity, which supplies the cutoff to the field. Einstein's equation then leads to a power-law expansion of the form

$$a(t) \sim \exp t^{1-p} \qquad (48)$$

where p < 1. This describes a universe with accelerated expansion, thus having dark energy. The equivalent cosmological constant decays in time like $t^{-2p}$, circumventing the usual "fine-tuning problem". Vortex activities in the superfluid creates quantum turbulence, in which all matter was created during a initial "inflation era". Many observed phenomena, such as dark mass halos around galaxies, can be explained.